\pdfoutput=1
\documentclass[a4paper,oneside,twocolumn,notitlepage]{extarticle_ecoc}
\usepackage{color}
\usepackage{amsmath}
\usepackage{graphicx}
\PassOptionsToPackage{normalem}{ulem}
\usepackage{ulem}

\makeatletter

\pdfpageheight\paperheight
\pdfpagewidth\paperwidth

\usepackage{tikz}
\usetikzlibrary{calc}

\usepackage{enumitem}

\usepackage{ecoc}
\usepackage[protrusion=true,expansion=true]{microtype}

\makeatother

\usepackage[style=ieee]{biblatex}
\usepackage{biblatex2bibitem}
\begin{document}

\title{A Sequence Selection Bound for the Capacity\\
of the Nonlinear Fiber Channel}

\author{Stella Civelli \textsuperscript{(1,2)}, Enrico Forestieri\textsuperscript{(1,2)},
Alexey Lotsmanov\textsuperscript{(3)}, Dmitry Razdoburdin\textsuperscript{(3,4)}
, Marco Secondini\textsuperscript{(1,2)}}

\maketitle
\begin{strip}
\begin{author_descr}

\textsuperscript{(1)} Tecip Institute, Scuola Superiore Sant'anna
\textcolor{blue}{\uline{stella.civelli@santannapisa.it}}

\textsuperscript{(2)} PNTLab, Consorzio nazionale interuniversitario
per le telecomunicazioni (CNIT)

\textsuperscript{(3)} Moscow Research Center, Huawei Technologies
Co., Ltd., Moscow, Russia

\textsuperscript{(4)} Sternberg Astronomical Institute, Moscow M.V.
Lomonosov State University, Moscow, Russia
\end{author_descr}

\end{strip}

\setstretch{1.1}

\renewcommand\footnotemark{} \renewcommand\footnoterule{} \let\thefootnote\relax\footnotetext{978-1-6654-3868-1/21/\$31.00 \textcopyright 2021 IEEE} 

\begin{strip}
\begin{ecoc_abstract}
A novel technique to optimize the input distribution and compute a
lower bound for the capacity of the nonlinear optical fiber channel
is proposed. The technique improves previous bounds obtained with
the additive white Gaussian noise decoding metric.
\end{ecoc_abstract}
\end{strip}

\section{Introduction}

The capacity of the fiber channel in the nonlinear regime is not known
\cite{secondini_JLT2017_scope,agrell17ecoc,SECONDINI2020867}, but
only upper-limited by the linear capacity \cite{Kramer2015} and lower-limited
by numerous bounds \cite{splett1993ultimate,mitra:nature,Essiambre:JLT0210,Mecozzi:JLT0612,dar:OL2014,secondini:ecoc17,garcia2020mismatched1,garciagomez2021mismatched}.
Most of the bounds are obtained by computing an achievable information
rate (AIR) with an optimized decoding metric, while considering a
simple fixed input distribution---typically, i.i.d. samples with
Gaussian distribution \cite{splett1993ultimate,mitra:nature,Mecozzi:JLT0612,dar:OL2014,secondini:ecoc17,garcia2020mismatched1,garciagomez2021mismatched}
or multiple rings with uniform phase \cite{Essiambre:JLT0210}.

In this work, we propose a novel sequence selection technique to optimize
the input distribution and compute an improved lower bound for the
capacity of the nonlinear fiber channel. The technique uses a rejection
sampling method to select only the sequences of symbols that generate
less nonlinear interference. The AIR achievable when encoding information
on the selected sequences is then lower bounded by removing the rate
loss caused by the selection procedure.

\section{Sequence selection}

\begin{figure}
\includegraphics[width=1\columnwidth]{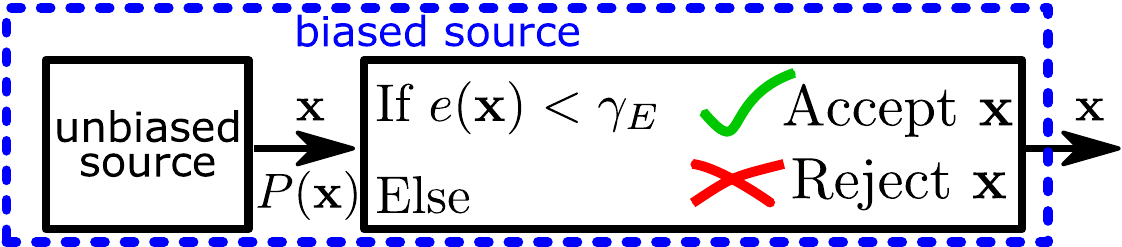}

\caption{\label{fig:biasedsource}The biased source, obtained with a selection
procedure from an unbiased source.}
\end{figure}

The sequence selection procedure---sketched in Fig.\ \ref{fig:biasedsource}---is
a sort of rejection sampling method, in which a random sequence of
$N$ symbols $\mathbf{x}$ is drawn from a given \emph{unbiased source}
with probability distribution $P(\mathbf{x})$, accepted if it meets
a certain condition, or rejected otherwise. The combination of the
unbiased source and the rejection method forms the \emph{biased} source,
which generates the symbols at the channel input. The unbiased source
can be arbitrarily selected, i.e., the $N$ symbols can be drawn from
a continuous (e.g., Gaussian) or discrete (e.g., quadrature amplitude
modulation (QAM)) constellation, and can be independent or correlated
(e.g., obtained as the output of a finite-block-length distribution
matcher that implements probabilistic shaping (PS). The acceptance
criterion is defined by selecting a proper metric $e(\cdot)$---which
measures the amount of nonlinear interference generated by a sequence---and
a threshold $\gamma_{E}$, so that a sequence is accepted only if
the metric $e(\mathbf{x)}$ is below the threshold, i.e., if it generates
a small amount of nonlinear interference. The probability of the sequences
generated by the biased source is 
\begin{equation}
P_{b}(\mathbf{x})=\begin{cases}
P(\mathbf{x})/\eta & e(\mathbf{x})<\gamma_{E}\\
0 & e(\mathbf{x})\geq\gamma_{E}
\end{cases}\label{eq:prob_biased}
\end{equation}
 where $\eta=\mathrm{Pr}\{e(\mathbf{x})<\gamma_{E}\}$ is the acceptance
probability, which can be approximated as $\eta\approx N_{s}/N_{t}$
where $N_{s}$ is the number of selected sequences and $N_{t}$ is
the number of tested sequences.

The selection metric $e(\cdot)$ can be defined and computed in different
ways.  In this work, we consider only intrachannel nonlinearity,
estimated by using the split-step Fourier method (SSFM). The metric
is defined as $e(\mathbf{x})=||\mathbf{x}-\mathbf{y}||$, where $\mathbf{x}$
is the transmitted sequence and $\mathbf{y}$ is the corresponding
received sequence after a single-channel noiseless propagation. The
sequences from the biased source are obtained by the following steps:
\begin{enumerate}
\item Draw $N_{t}$ test sequences $\{\mathbf{x}_{k}\}$, with $k=1,\dots,N_{t}$,
from the unbiased source.
\item Form the sequence $\mathbf{s}=(\mathbf{x}_{1},\dots,\mathbf{x}_{N_{t}})$
of length $N_{t}N$, obtained by concatenating all the test sequences.
\item Run a single-channel noiseless simulation with input $\mathbf{s}$,
including all the steps that will be included in the system except
for digital backpropagation (DBP)---i.e., modulation, SSFM propagation,
dispersion compensation, matched filtering and sampling, mean phase
rotation compensation---to obtain the corresponding received sequence
$\mathbf{r}=(\mathbf{y}_{1},\dots,\mathbf{y}_{N_{t}})$.
\item Accept only the $N_{s}$ sequences with $e(\mathbf{x}_{k})=||\mathbf{x}_{k}-\mathbf{y}_{k}||<\gamma_{E}$.
\end{enumerate}
As far as it concerns the system performance, we expect it to improve
as $\gamma_{E}$ becomes smaller---i.e., as the maximum amount of
nonlinear interference that can be generated by each sequence diminishes---at
least in the same scenario considered for sequence optimization (single-channel,
dispersion compensation only). However, by reducing $\gamma_{E}$
also $\eta$ decreases, meaning that less sequences are available
to encode information. This \emph{rate loss} is accounted for in the
computation of the AIR. When the transmitted symbols are drawn from
a Gaussian constellation, we consider the AIR with symbol-wise decoding,
$\mathrm{AIR_{s}}$. When sequence selection is not applied, $\mathrm{AIR_{s}}$
is evaluated as in \cite{secondini2019JLT}, assuming a mismatched
decoding metric optimized for the additive white Gaussian noise (AWGN)
channel. Conversely, when sequence selection is applied, the following
lower bound holds
\begin{equation}
\mathrm{AIR_{s}}\geq\mathrm{AIR_{s}^{(u)}}+\frac{\log_{2}\eta}{2N}\text{ in bits}/\text{symbol/pol}\label{eq:MIbiased}
\end{equation}
where $\mathrm{AIR_{s}^{(u)}}$ is evaluated assuming that the received
sequence $\mathbf{x}$ has unbiased probability $P(\mathbf{x})$,
i.e., transmitting the sequences obtained from the biased source but
computing the AIR with the same expression as in the case without
sequence selection. The inequality in (\ref{eq:MIbiased}) is obtained
using $P_{b}(\mathbf{x})\leq\eta P(\mathbf{x})$, which follows from
(\ref{eq:prob_biased}) and implies a loss of at most $\log_{2}(1/\eta)$
bits on a sequence of $N$ 2-polarization symbols.

On the other hand, when the transmitted symbols are drawn from a QAM
constellation, we consider the AIR with bit-wise decoding, $\mathrm{AIR_{b}}$,
still with the same mismatched AWGN decoding metric. When sequence
selection is not applied, $\mathrm{AIR_{b}}$ is evaluated as in \cite{fehenberger2016JLT,alvarado2018achievable}.
When sequence selection is applied, a lower bound analogous to (\ref{eq:MIbiased})
holds. 

\section{System setup and results}

\begin{figure}
\includegraphics[width=1\columnwidth]{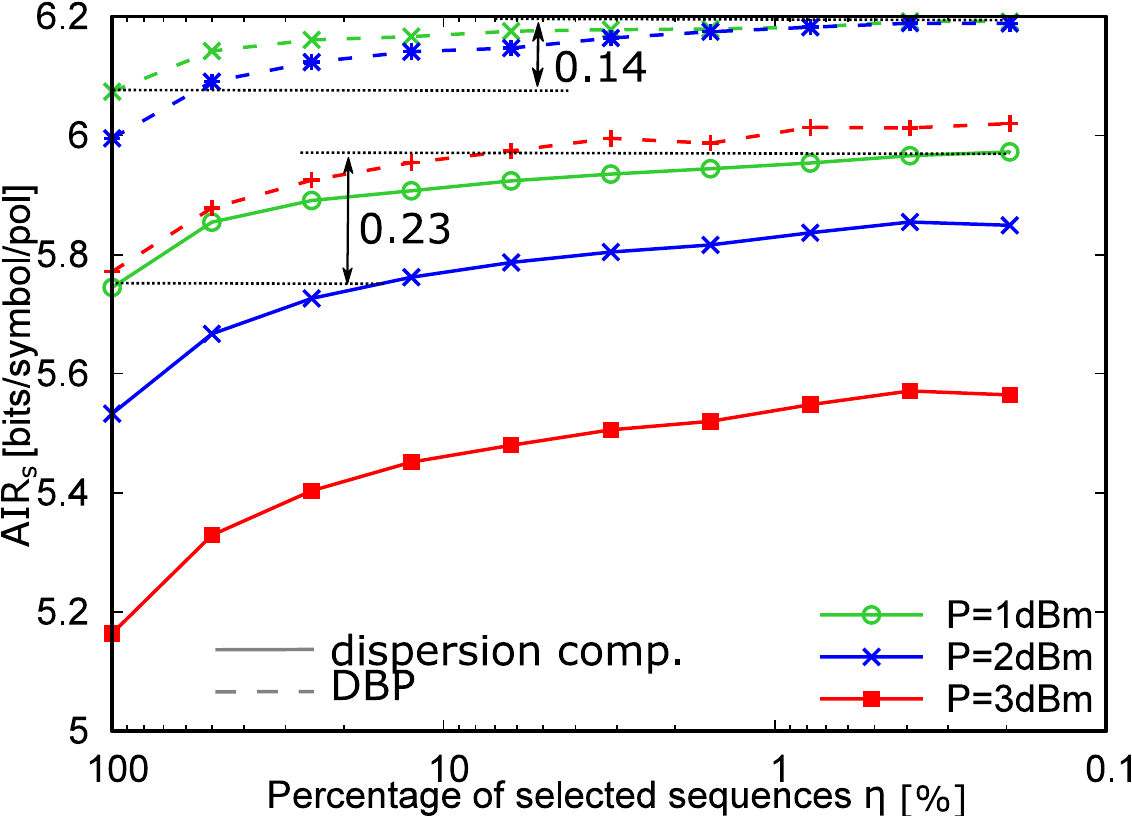}

\caption{\label{fig:100km}$\mathrm{AIR_{s}}$ vs selection rate for the EDFA
link and i.i.d. Gaussian symbols as unbiased source.}
\end{figure}
The system setup is the same considered in \cite{secondini2019JLT}.
A dual polarization WDM signal made of $5$ $R_{s}=50$\,GBd channels,
with $50$\,GHz spacing, sinc pulses, and Gaussian or QAM symbols
with PS, is launched into a $1000$\,km link. The link is made of
$10\times100$km spans of single mode fiber ($D=17$\,ps/nm/km, $\gamma=1.3$~W$^{-1}$/km,
and $\alpha_{\text{dB}}=0.2$\,dB/km), after each span an erbium-doped
fiber amplifier (EDFA) with spontaneous emission coefficient equal
to $1$ compensates for loss. The ideal Raman amplification (IDRA)
case is also considered. At the RX, the central channel is demultiplexed,
DBP or dispersion compensation is applied, followed by matched filter
and sampling at symbol time $1/R_{s}$. After a mean phase rotation
removal, the lower bound to the $\mathrm{AIR_{s}}$ or $\mathrm{AIR_{b}}$
is evaluated. The PS is implemented through the probabilistic amplitude
shaping (PAS) approach \cite{bocherer2015bandwidth}, using either
i.i.d. symbols drawn from a Maxwell-Boltzmann (MB) distribution---optimal
in the linear regime---or the enumerative sphere shaping (ESS) \cite{gultekin2018Sphereshaping}
with optimized block length equal to $256$. The length of the sequences
is $N=256$ $2$-pol QAM symbols. The number of tested sequences is
$N_{t}=2^{16}$. The selection procedure to determine which sequences
are accepted (used in simulations to compute the AIR) or rejected
is performed in a noiseless single-channel scenario at a launch power
corresponding to the optimal power for the case without sequence selection.

Fig.~\ref{fig:100km} shows the lower bound (\ref{eq:MIbiased})
as a function of the selection rate $\eta$ for the EDFA link and
different launch powers, both without DBP (solid lines) and with ideal
single-channel DBP (dashed lines). The optimal power is $P=1$dBm
at $\eta=100\%$. The figure shows that the performance improves as
$\eta$ decreases, though it seems to approach a maximum near the
smallest value of $\eta$ considered in the figure (which is limited
by the number of tested sequences). This happens because (i) the sequences
have been optimized for the single channel scenario without DBP, and
not for the consider scenario, making the sequences not optimal, and
(ii) the loss due to sequence selection---the second term in (\ref{eq:MIbiased})---increases
when $\eta$ decreases, so that eventually all the curve must decrease
again and vanish when $\eta\rightarrow0$. Overall, the maximum gain
in the case without DBP is $0.23$bits/symbol/pol obtained with $\eta=0.19\%$,
and $0.14$bits/symbol/pol obtained with $\eta=0.39\%$ for the case
with DBP. The gain obtained with DBP---though smaller---is particularly
interesting since the sequences have been selected in the single channel
scenario without DBP, that is, to minimize the intrachannel nonlinearity,
which is instead fully compensated for by DBP in this case. This means
that the same sequences that are ``good'' for intrachannel nonlinearity
are good also for mitigating interchannel nonlinearities. However,
we expect to achieve higher gains by employing a selection metric
that measures also interchannel nonlinearity.

\begin{figure}
\includegraphics[width=1\columnwidth]{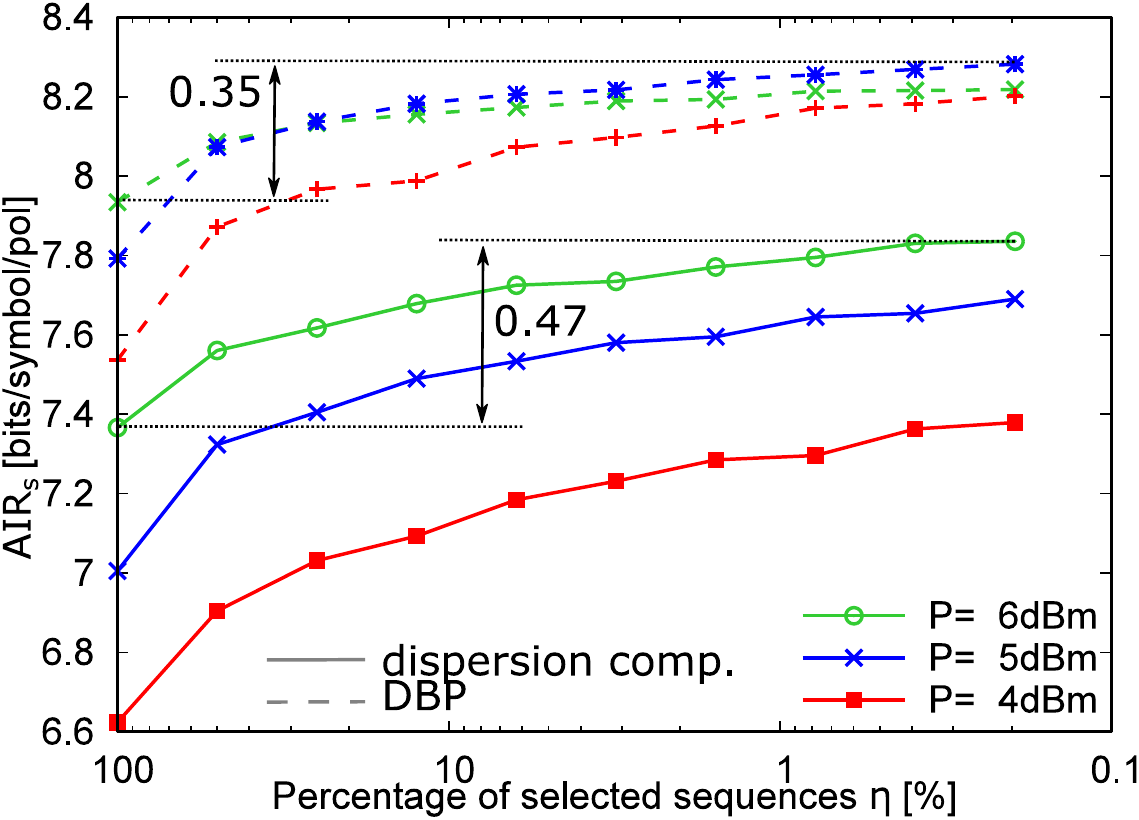}

\caption{\label{fig:Raman}$\mathrm{AIR_{s}}$ vs selection rate for the IDRA
link and i.i.d. Gaussian symbols as unbiased source.}
\end{figure}

Fig.\ \ref{fig:Raman} shows the results for the IDRA link. The qualitative
behaviour is the same as in the EDFA link of Fig.~\ref{fig:100km},
but with larger gains: the maximum gains are $0.47$\,bits/symbol/pol
and $0.35$\,bits/symbol/pol without and with DBP, respectively.
The reason why the gains are higher in the IDRA link than in the EDFA
link is under investigation, but we note that the same behaviour is
observed in the case studied in \cite{secondini2019JLT}, where it
can be explained by the higher coherence (in time and frequency) of
the cross-phase modulation term in the IDA link. In \cite{secondini2019JLT},
however, the AIR gains are obtained by optimizing the decoding metric
rather than the input distribution. Moreover, we expect that by combining
the sequence selection approach proposed here for the optimization
of the input distribution with the optimized decoding metric employed
in \cite{secondini2019JLT,garciagomez2021mismatched}, we might further
improve the AIR (and capacity bounds).

\begin{figure}
\includegraphics[width=1\columnwidth]{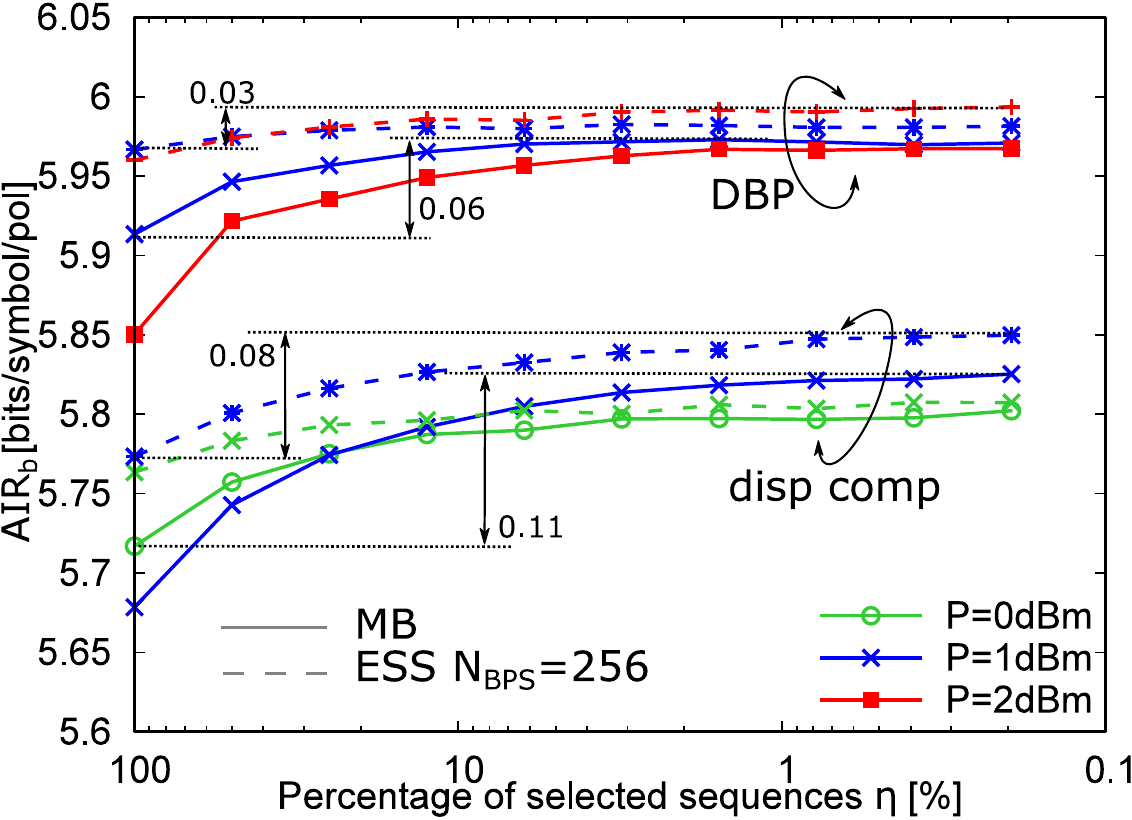}

\caption{\label{fig:100kmPAS}$\mathrm{AIR_{b}}$ vs selection rate for the
EDFA link and PAS-$256$\,QAM symbols with i.i.d. MB distribution
(solid lines) or drawn by ESS with optimized block length (dashed
lines) as unbiased source.}
\end{figure}

Finally, we test the proposed approach for a discrete constellation.
Fig.~\ref{fig:100kmPAS} shows the performance of sequence selection
when a PAS-$256$\,QAM constellation with rate $R=6.4$\,bits/symbol/pol
is used, with or without DBP. When i.i.d. symbols with the MB distribution
are considered as a starting point ($\eta=100\%)$ for the unbiased
distribution (solid lines), the performance improves by $0.11$bits/symbol/pol
and $0.06$bits/symbol/pol, without and with DBP, respectively. Interestingly,
the gain provided by sequence selection in the case without DBP is
$0.05$\,bits/symbol/pol larger than the gain provided by PAS with
optimized block length \cite{geller2016shaping,civelli2020interplayECOC,fehenberger2020mitigating}.
When DBP is used, the two techniques---sequence selection with MB
symbols and optimized ESS without sequence selection---provide the
same gain. Finally, the highest AIR is obtained when considering
ESS with optimized block length \cite{civelli2020interplayECOC} as
a starting point ($\eta=100\%)$ for the unbiased distribution (dashed
lines). In this case, sequence selection yields a gain of $0.08$\,bits/symbol/pol
and $0.03$\,bits/symbol/pol without and with DBP, respectively.

\section{Conclusions}

We have proposed a novel sequence selection technique to lower-bound
the capacity of the nonlinear optical fiber channel. Using a simple
numerical optimization, the proposed technique improves the AIR obtained
with an AWGN decoding metric, with significant gains over both EDFA
and IDRA links, with both continuous and discrete constellations.
The use of a more accurate selection metric (e.g., accounting also
for interchannel nonlinearity) and the combination with an improved
decoding metric \cite{secondini2019JLT,garciagomez2021mismatched}
are expected to further increase the lower bounds provided in this
work.

\section{Acknowledgement}

This work was funded in part by Huawei.

\printbibliography
\end{document}